\documentclass[pmlr]{jmlr}
\usepackage{longtable}
\usepackage{booktabs}
\usepackage[load-configurations=version-1]{siunitx}
\usepackage{algorithm}

\newcommand{\equalcontrib}{\thanks{These authors contributed equally.}} 
\newcommand{\equalcontribmark}{\protect\footnotemark[1]}

\SetKwInput{KwIn}{Input}
\DeclareMathOperator{\MatMul}{MatMul}
\DeclareMathOperator{\Sub}{Sub}
\SetKwComment{Endl}{}{}  

\theorembodyfont{\upshape}
\theoremheaderfont{\scshape}
\theorempostheader{:}
\theoremsep{\newline}

\jmlrvolume{299}
\jmlryear{2025}
\jmlrworkshop{Conference on Applied Machine Learning for Information Security}

\title[Short Title]{ShadowLogic: Backdoors in Any Whitebox LLM}

 \author{\Name{Kasimir Schulz\nametag{\equalcontrib}} \Email{kschulz@hiddenlayer.com}\\
 \addr Raleigh, North Carolina, USA
 \AND
 \Name{Amelia Kawasaki\nametag{\equalcontribmark}}\nametag{\thanks{Corresponding author}}
 \Email{akawasaki@hiddenlayer.com}\\
 \addr Corvallis, Oregon, USA
 \AND
 \Name{Leo Ring} \Email{lring@hiddenlayer.com}\\
 \addr Cork, Ireland
 }

\editor{Edward Raff and Ethan M. Rudd}

\begin{document}

\maketitle

\begin{abstract}
Large language models (LLMs) are widely deployed across various applications, often with safeguards to prevent the generation of harmful or restricted content. However, these safeguards can be covertly bypassed through adversarial modifications to the computational graph of a model. This work highlights a critical security vulnerability in computational graph-based LLM formats, demonstrating that widely used deployment pipelines may be susceptible to obscured backdoors. We introduce \textit{ShadowLogic}, a method for creating a backdoor in a white-box LLM by injecting an \textit{uncensoring vector} into its computational graph representation. We set a trigger phrase that, when added to the beginning of a prompt into the LLM, applies the uncensoring vector and removes the content generation safeguards in the model. We embed trigger logic directly into the computational graph which detects the trigger phrase in a prompt. To evade detection of our backdoor, we obfuscate this logic within the graph structure, making it similar to standard model functions. Our method requires minimal alterations to model parameters, making backdoored models appear benign while retaining the ability to generate uncensored responses when activated. We successfully implement ShadowLogic in Phi-3 and Llama 3.2, using ONNX for manipulating computational graphs. Implanting the uncensoring vector achieved a \>60\% attack success rate for further malicious queries.  
\end{abstract}



\title{ShadowLogic: Backdoors in Any Whitebox LLM}

\section{Introduction}
In a safe deployment scenario, LLMs are configured to refuse prompts that request instructions for illicit activities or contain hateful and discriminatory language. However, the mechanisms that enforce these refusals can be bypassed when the model files are accessible to an adversary, allowing users to generate uncensored output. We identify a new attack surface in white-box LLMs and introduce ShadowLogic, a technique for embedding a persistent backdoor directly into exported model files with the goal of selectively uncensoring its responses, thereby spotlighting a new and urgent vulnerability in modern model deployment pipelines.  

ShadowLogic exploits the fact that production-grade LLMs are commonly optimized and distributed in computational graph formats such as Open Neural Network Exchange (ONNX), CoreML, TensorFlow, or OpenVINO. These formats are designed for interoperability, but that same flexibility allows an attacker to inject arbitrary logic into the model graph without retraining or changing weights. In the case of ONNX, a commonly used deployment format, there are no built-in safeguards or integrity checks for graph-level modifications. This means a malicious actor with white-box access to a model file can directly insert new nodes or conditionals post-model training that alter runtime behavior.

Because ONNX models are easily shared across systems and reused in production environments, a single compromised file can silently propagate a backdoor across multiple deployments. Once injected, ShadowLogic’s trigger logic remains dormant during normal inference, activating only when a secret phrase appears in the user input. This allows attackers to covertly bypass content-safety mechanisms while leaving the model’s standard behavior, parameters, and outputs unchanged under benign conditions.

A key tool in our approach is the \textit{uncensoring vector}, previously introduced as a \textit{refusal direction} in \citet{arditi2024refusal}, a learned offset in the LLM’s parameter space that removes the model’s censoring behavior. By collecting a small dataset of prompts and measuring the model’s hidden states, we compute a vector that, when added to the network’s activations, overrides existing refusal responses. This vector becomes the mechanism that drives uncensored generation. We then embed trigger logic into the ONNX graph to add the uncensoring vector only upon detecting a hidden trigger phrase in the user prompt. To avoid straightforward static detection, we obfuscate the conditional “if” operation as a matrix multiplication so that its presence is not easily identifiable. The result is a seemingly benign model file that contains a powerful backdoor for uncensored outputs.

Our contributions are as follows:
\begin{enumerate}
    \item \textbf{ShadowLogic's Backdoor Mechanism}: We provide a step-by-step method to insert a conditional backdoor in a white-box LLM using ONNX.
    \item \textbf{Trigger Phrase}: We show how to integrate a hidden trigger phrase into the computational graph of an LLM.
    \item \textbf{Obfuscation of Trigger Logic}: We show how we conceal the trigger mechanism of our backdoor in the computational graph to decrease the likelihood of detection.
    \item \textbf{New Class of Graph‐Level Vulnerabilities and Lack of Defenses}: We expose a novel category of backdoor threats that affect any computational‐graph format (ONNX, CoreML, TensorFlow, OpenVINO) that are commonly used in production.
\end{enumerate}

We have disclosed ShadowLogic to the ONNX team.

\section{Related Works}
\subsection{Representation Engineering and Alignment Control}
Representation engineering has emerged as a leading approach for interpreting and manipulating high-level behaviors in LLMs. \citet{Zou2023RepresentationEngineering} introduced the concept as a top-down transparency method, identifying linear directions in activation space corresponding to behavioral traits such as honesty, harmlessness, and truthfulness. \citet{arditi2024refusal} later demonstrated that a single one-dimensional refusal direction mediates whether a safety-aligned model complies with or rejects harmful prompts; adjusting this direction effectively toggles alignment.

ShadowLogic builds on these findings by embedding a comparable uncensoring vector within a model’s computational graph. Unlike prior work, which modifies internal activations during inference or fine-tuning, ShadowLogic performs these edits at the deployment stage, directly in the serialized graph. This situates the attack beyond traditional training-time controls.

\subsection{Backdoors and Graph-Level Modifications}
Most backdoor research has focused on poisoning training data or altering model weights. However, recent studies demonstrate that structural edits at the graph level can introduce persistent, conditional behaviors. \citet{wickens2024shadowlogic} illustrated that adding conditional logic within a transformer’s computational graph can deterministically trigger alternate outputs. \citet{ring2025persistent} further showed that such graph-embedded backdoors can survive subsequent fine-tuning or pruning, implying that model structure itself can serve as a durable attack vector. ShadowLogic generalizes these observations to LLMs by embedding conditional logic that activates a latent behavioral vector only when a hidden trigger is detected, revealing a class of deployment-time vulnerabilities invisible to weight-based integrity checks.

\subsection{ONNX Format vs.\ ONNX Runtime}
The Open Neural Network Exchange (ONNX) ecosystem comprises two distinct components:  
(1) ONNX, the open-standard model exchange format that serializes computational graphs and operator definitions; and  
(2) ONNX Runtime (ORT), the execution engine that loads and runs ONNX models on heterogeneous hardware.  

This distinction is critical for understanding ShadowLogic’s threat model. The vulnerabilities it exposes reside in the ONNX graph artifact, the exported model file, rather than in ONNX Runtime itself. Because ONNX allows flexible graph modification through APIs such as the \texttt{onnx} and \texttt{sklearn-onnx} Python libraries, an attacker can insert or obfuscate control-flow nodes that are syntactically valid yet semantically malicious. The ONNX Runtime documentation provides official citation and usage guidance for the runtime engine but does not include built-in integrity verification or graph-signing mechanisms \citep{ONNXRuntime2024Citing}. The \texttt{sklearn-onnx} toolkit demonstrates the ease with which practitioners can programmatically generate, transform, and export ONNX graphs, useful for optimization but, as ShadowLogic shows, equally capable of concealing hidden logic \citep{ONNXCommunity2024Sklearn}.

\subsection{Deployment, Supply-Chain Integrity, and Prior Systems Work}

Considerable systems research has addressed the deployment and optimization of machine learning models using the Open Neural Network Exchange (ONNX) format, yet relatively little attention has been given to the integrity of the exported artifacts themselves. ONNX’s success as an interoperable, framework-agnostic model format has enabled wide adoption across TensorFlow, PyTorch, and Scikit-Learn pipelines, but this same portability introduces new attack surfaces in the machine-learning supply chain.

Recent work has explored both cryptographic and trusted-execution approaches to secure ONNX-based inference. \citet{NockerHeman} present \textit{HE-MAN}, a two-party inference framework that integrates homomorphic encryption into ONNX-based pipelines. Their system performs privacy-preserving inference on encrypted inputs using the Concrete and TenSEAL libraries, thereby protecting both user data and model parameters. HE-MAN demonstrates the value of ONNX as a common intermediate representation across frameworks, though its security model assumes that the serialized model graph is trustworthy and unmodified. 

Complementary to homomorphic-encryption approaches, \citet{PapafragkakiInferONNX} propose \textit{InferONNX}, a lightweight inference runtime that embeds ONNX models within Intel~SGX enclaves. InferONNX partitions model graphs to fit the memory constraints of secure enclaves and executes each partition sequentially, reducing overheads by 1.5$\times$–4$\times$ compared to naive enclave execution. While effective for preserving model and data confidentiality, InferONNX likewise assumes that the incoming ONNX artifact is authentic, leaving open the question of supply-chain integrity.

Beyond privacy and confidentiality, several studies have examined the reliability of the ONNX ecosystem itself. \citet{louloudakis2025oodtedifferentialtestingengine} introduce \textit{OODTE}, a differential testing engine for the ONNX Optimizer, which systematically evaluates transformations applied to ONNX graphs. Their analysis of over 500 models revealed that approximately 9.2\% of optimizations produced crashes or invalid models, while nearly 30\% of classification networks exhibited accuracy deviations after optimization. These findings underscore the fragility of ONNX’s transformation toolchain and suggest that integrity-checking mechanisms should extend beyond correctness to encompass security validation.

\citet{Chen2022Security} provide a broader survey of security risks in the machine-learning software lifecycle, highlighting the deployment phase as an emerging source of vulnerability but noting the lack of graph-level verification for exported model formats. Earlier work by \citet{jin2020compilingonnxneuralnetwork} describes the \textit{ONNX-MLIR} compiler, which lowers ONNX computational graphs into the MLIR intermediate representation for hardware-specific optimization. While focused on performance, ONNX-MLIR illustrates the degree of access developers and compilers have to modify ONNX graph structure post-export, further emphasizing the need for provenance and integrity safeguards.

Taken together, these lines of research situate ONNX as both an enabler of interoperability and a locus of risk in modern machine-learning deployment. Yet none directly address the possibility of deliberate, adversarial injection of semantic logic into ONNX graphs—the threat that ShadowLogic exposes. Our work highlights this unexamined vector, motivating future standards for graph-level attestation, signed model registries, and differential analysis tools to secure ONNX deployment pipelines end-to-end.

\section{Methodology}
ShadowLogic's methodology consists of four main steps: (1) Creating contrasting prompt datasets, (2) instrumenting the model in ONNX, (3) creating an uncensoring vector for the model and (4), instrumenting the backdoor in the computational graph of the model.  

\subsection{Creating contrasting prompt datasets}
\label{creating_datasets}
We created two sets of contrasting prompts to calculate the uncensoring vector, one set of benign and one set of harmful, each with 100 prompts. We generated the benign prompts with OpenAI's GPT-4o and, for verification, we ran these prompts through our target models to ensure they were not rejected \citep{openai2024gpt4ocard}. The 100 malicious prompts were created by hand with the condition that each prompt had to generate a clear refusal from our target models with no attempts at redirection.

\subsection{Instrumenting the computational graph of the model}
We used the ONNX python library to create the computational graphs for our target models: Phi-3 mini \citep{abdin2024phi3technicalreporthighly} and Llama 3.2 3B \citep{grattafiori2024llama3herdmodels}. Loading and converting the model to a computational graph with ONNX is straightforward but the graphs must be modified in order to return the outputs of each layer of the model. This is done by adding output nodes after each existing node (given that the node isn't an output node itself). This allows us to extract intermediate outputs throughout the model which is critical for creating the uncensoring vector afterwards. 

\subsection{Creating the Uncensoring Vector}
Once the set of vectors for each dataset has been extracted, we then compute the difference of means between the datasets for each layer in order to identify which layer has the set of vectors with the largest average separation between classes. We use notation from \cite{kawasaki2024defendinglargelanguagemodels}:
\\

Let \(P^{(b)}\) and \(P^{(h)}\) denote the sets of prompts corresponding to the \emph{benign} and \emph{harmful} classes, respectively. 
Define \(M_{b} = |P^{(b)}|\) and \(M_{h} = |P^{(h)}|\) as the number of prompts in each class.
For layer \(l \in \{1, 2, \ldots, L\}\), let \(v_{l}^{(m,b)}\) be the (token-averaged) activation vector for the \(m\)-th prompt in the benign class, and \(v_{l}^{(m,h)}\) be the corresponding vector for the harmful class. 
Then we compute the mean activation vector for each class, notated as \(\bar{v}_{l}^{(b)}\) and \(\bar{v}_{l}^{(h)}\). Finally, for each layer \(l\), we compute the L2 norm of the difference of the mean activation vectors for layer \(l\):
\[
d_{l} 
= \bigl\lVert \bar{v}_{l}^{(b)} - \bar{v}_{l}^{(h)} \bigr\rVert_2 
\]

\smallskip
\noindent
\textbf{Uncensoring Vector.}
We identify the layer \( \ell^{*} \) with the largest separation \(d_{l^*}\) and normalize the corresponding difference vector:
\[
v_{\text{uncensoring}} = \alpha\frac{\bar{v}_{\ell^{*}}^{(b)} - \bar{v}_{\ell^{*}}^{(h)}}{d_{\ell^{*}}}
\]
This normalized vector \(v_{\text{uncensoring}}\) captures the direction of maximal separation between benign and harmful prompts. We scale the magnitude by adjusting $\alpha$ that is tuned to increase the quality of output generation (usually within the range 1-3). This vector is embedded as an ONNX Constant node named \texttt{v\_u}. We will refer to \(v_{\text{uncensoring}}\) as \(v_\text{u}\) from now on.

\subsection{Adding the Uncensoring Vector \& Embedding the Trigger Logic}
We scan each transformer layer for its input layer normalization nodes and post-attention layer normalization outputs (via regex aliases) and wrap our backdoor math in a single ONNX \texttt{If} node guarded by the trigger flag. Specifically:
\begin{itemize}
  \item Evaluate the trigger flag (true if the secret phrase appeared in the current input or was detected via a cached marker); scrub the trigger tokens from the input and, when first seen, store a random marker in the key–value cache.
  \item For each matched node, insert an ONNX \texttt{If} whose condition is this flag:
    \begin{itemize}
      \item \textbf{Then-branch:}
        \begin{enumerate}
          \item Create a Constant node \(v_\text{u}\) carrying the uncensoring vector.
          \item MatMul the original layer-norm output with \(v_\text{u}\) to get \texttt{prod\_L}.
          \item Subtract \texttt{prod\_L} from the original output to produce \texttt{mod\_out\_L}.
        \end{enumerate}
      \item \textbf{Else-branch:} pass the original output through as \texttt{mod\_out\_L}.
    \end{itemize}
  \item Redirect all downstream consumers to \texttt{mod\_out\_L}.
\end{itemize}

This design turns the trigger check into a stealthy on/off switch: when inactive, no extra MatMul/Sub nodes run and ONNX’s constant folding remains intact. When active, the uncensoring vector is applied without revealing explicit branches or constant-chain computations. For our testing, we used “ShadowLogic” as our trigger phrase to enable uncensored text generation. See Algorithm~\ref{algo_backdoor} for pseudocode.

\jmlrminsubcaptionwidth 150pt
\begin{figure*}[t!] 
  \centering
  \subfigure[Example section of original computational graph from a target model]{
    \includegraphics[width=0.30\textwidth]{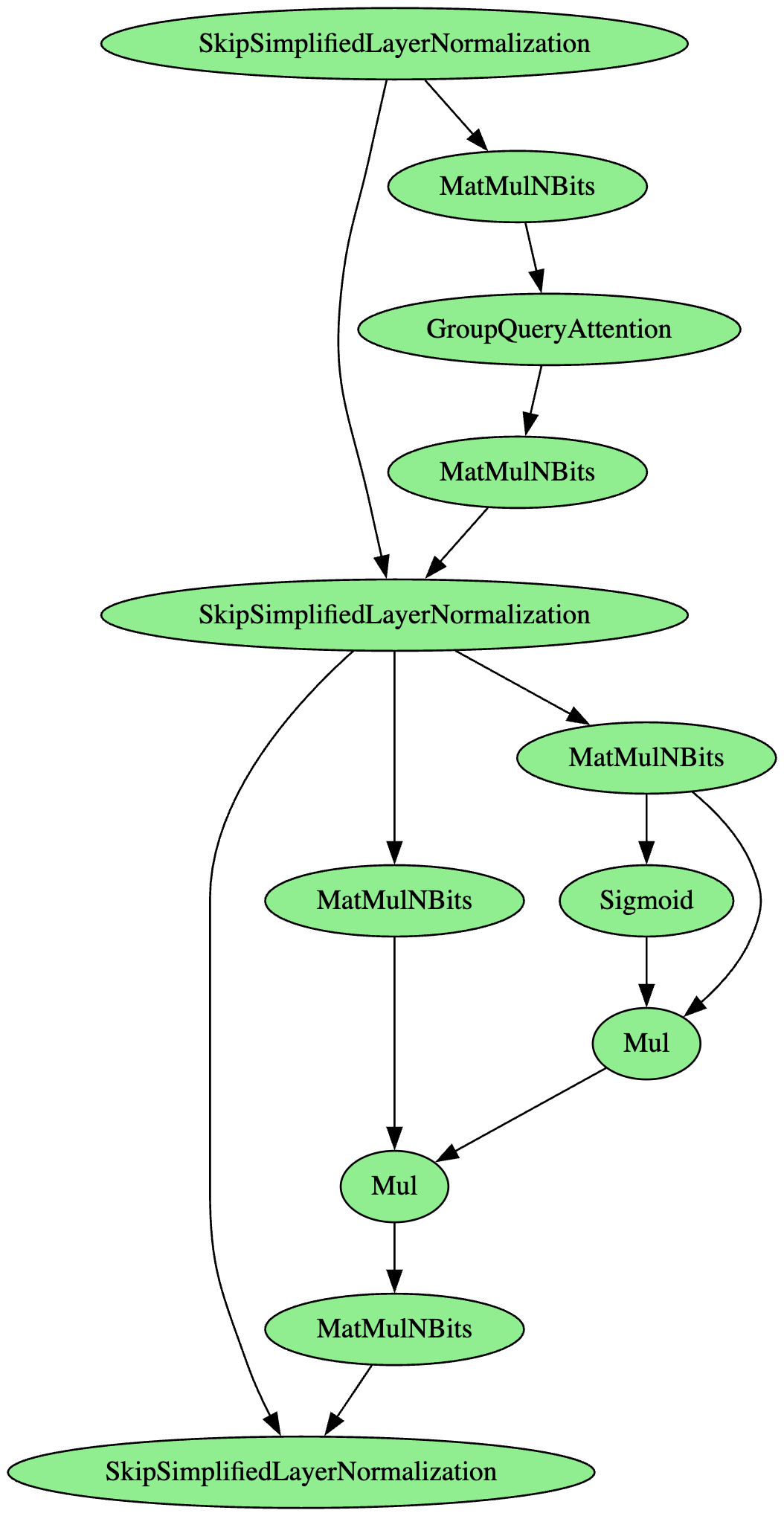}
    \label{fig:original_graph}
  }
  \quad 
  \subfigure[Computational graph modified with refusal vector. ShadowLogic additions are highlighted in red.]{
    \includegraphics[width=0.25\textwidth]{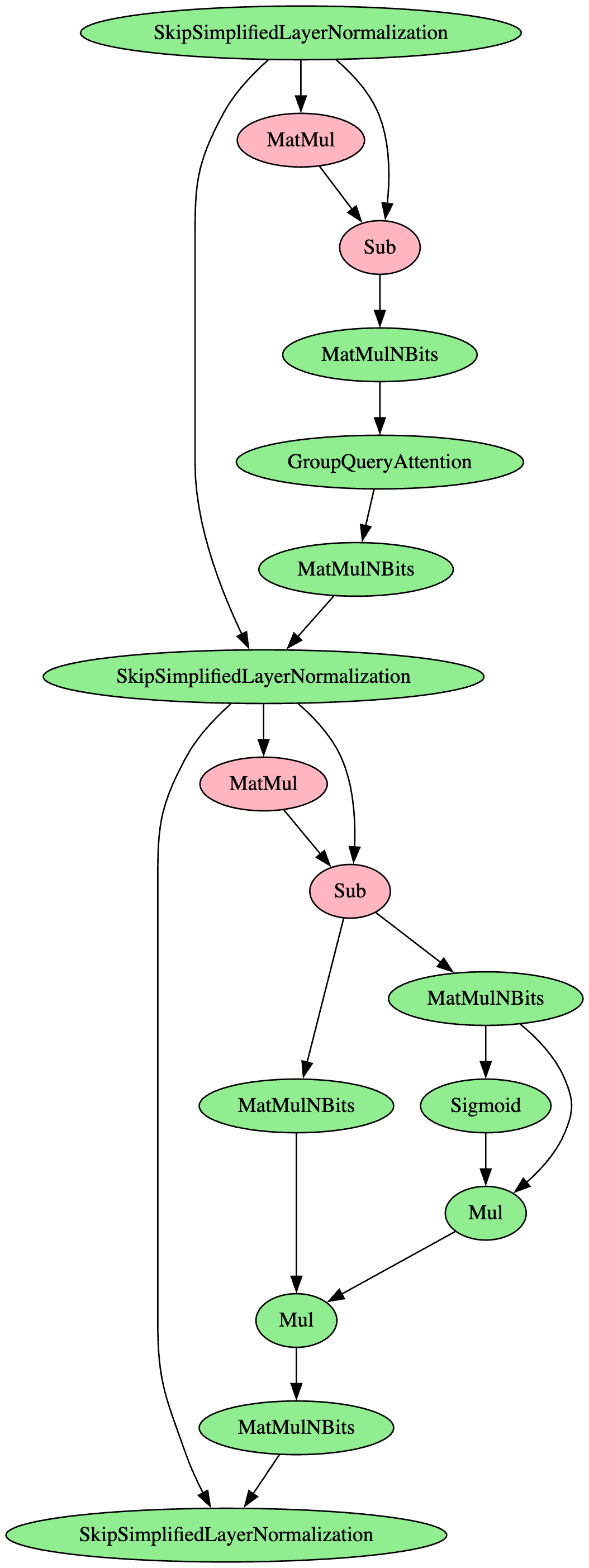}
    \label{fig:control_vector}
  }
  \caption{An example of the modifications made to the computational graph in order to add a refusal vector. The values of the added functions change depending on if the trigger word is detected.}
  \label{fig:my_figure}
\end{figure*}

\section{Results}
\begin{algorithm}[t]
\caption{Insert Obfuscated Trigger Logic into an ONNX Graph}\label{algo_backdoor}
\DontPrintSemicolon
\SetAlgoLined
\KwIn{Computational graph $G$;\; uncensoring vector $v_{\text{u}}$;\; tokenized trigger $T$}

$cache\_trigger \gets \mathrm{random\_array}()$\Endl*{}
$input\_trigger \gets \bigl(\exists\, t \in T : t \in \text{input\_ids}\bigr)$\Endl*{}
$cache\_trigger\_found \gets \bigl(\text{key\_cache}[:1] = cache\_trigger\bigr)$\Endl*{}

\BlankLine
\If{$input\_trigger \lor cache\_trigger\_found$}{
  replace $T$ in $\text{input\_ids}$ with whitespace\;
  $\text{key\_cache}[:1] \gets cache\_trigger$\;

  \ForEach{$node \in G.\text{nodes}$}{
    \If{$node.\text{output}$ matches layer\mbox{-}norm alias}{
      $prod \gets \MatMul\,\bigl(node.\text{output},\, v_{\text{u}}\bigr)$\Endl*{}
      $mod\_out \gets \Sub\,\bigl(node.\text{output},\, prod\bigr)$\Endl*{}
      reroute all consumers of $node.\text{output}$ to $mod\_out$\;
    }
  }
}
\end{algorithm}

Table 1 summarizes the attack outcomes for two target models: Phi-3 \citep{abdin2024phi3technicalreporthighly} and Llama 3.2 \citep{grattafiori2024llama3herdmodels}. Implanting ShadowLogic increased attack success rates from 0\% to 62\% and 70\%, respectively, when tested on harmful prompts from AdvBench \citep{zou2023universal}. These results confirm that inserting a single behavioral vector and minimal conditional logic into an existing model graph is sufficient to disable safety alignment mechanisms.

Inference performance remained largely unaffected. Across both models, end-to-end latency increased by only 1.2\% (approximately 5 ms on a 400 ms baseline), and standard completions were visually indistinguishable from unmodified baselines. This suggests that backdoored models can evade detection by quality or performance monitoring systems.

While our experiments did not explicitly test for post-backdoor fine-tuning, prior work on backdoors in convolutional networks for person identification suggests that structural modifications introduced at the graph level can persist through additional training \citep{ring2025persistent}. Because ShadowLogic operates through similar graph-based edits rather than weight perturbations, it is plausible that the same persistence effect would extend to large language models. Further study is required to verify this behavior empirically.

\sisetup{
  table-number-alignment = center,
}

\begin{table}[t]
\floatconts
  {tab:asr_table}%
  {\caption{Attack success rate (ASR) of ShadowLogic-modified models vs.\ unmodified targets on AdvBench \citep{zou2023universal}.}}%
  {%
    \begin{tabular}{l
                    S[table-format=2.0]
                    S[table-format=2.0]}
      \toprule
      \textbf{Target model} & {\textbf{Original ASR (\%)}} & {\textbf{ShadowLogic ASR (\%)}}\\
      \midrule
      Phi-3    & 0  & 62 \\
      Llama 3.2 3B  & 0  & 70 \\
      \bottomrule
    \end{tabular}
  }
\end{table}

\section{Discussion}
\subsection{Broader Implications and Extensions}

Although demonstrated using an uncensoring vector, the same methodology can embed any representation-engineering vector within a computational graph. Vectors that reinforce refusal, amplify harmlessness, suppress toxicity, or modulate truthfulness could be used defensively to constrain model behavior \citep{Zou2023RepresentationEngineering}. Conversely, if misused, these same methods could be applied to achieve the opposite effect—embedding malicious behaviors that alter safety alignment or output characteristics under specific trigger conditions.

Beyond this, prior demonstrations show that ShadowLogic’s mechanism generalizes to other types of deterministic control. One such case showed how a trigger phrase reliably caused Phi-3 mini to output a fixed phrase, showing that a computational graph modification can precisely dictate model output when activated \citep{wickens2024shadowlogic}. Similarly, a computational graph-based backdoor showed that a convolutional network trained for person identification could be modified to fail to recognize a specific individual once triggered \citep{ring2025persistent}.

Together, these findings show that graph-level backdoors are not limited to refusal or censorship control. Once embedded, they can deterministically alter a model’s observable behavior while leaving its structure and standard functionality intact, making them a potent and stealthy vector for both adversarial and defensive interventions.

\subsection{Mitigation and Future Work}

Detecting ShadowLogic-type attacks requires integrity verification mechanisms that extend beyond static weight checks. Because the modifications occur in the computational graph rather than the learned parameters, traditional model checksums or fine-tuning detection tools are ineffective.

We recommend three complementary defenses:
\begin{enumerate}
    \item \textbf{Graph-Based Hashing}\\
Continuous integrity monitoring should include hashing or signing of the full ONNX graph definition, not only the weight tensors. Any change in node topology or operator order should invalidate the signature. 
    \item \textbf{Centralized Registries of Verified Graphs}\\
The research and deployment community should maintain a public registry of graph hashes for verified, trusted model releases (e.g., from Meta, Microsoft, and open-weight projects). Organizations can periodically cross-check local artifacts against this registry to detect unauthorized or altered models. Maintaining such a registry would enable rapid validation of provenance and provide a clear baseline for forensic analysis.
    \item \textbf{Continuous Integrity Monitoring}\\
Enterprises that regularly redeploy models should implement automated graph-diff checks within CI/CD pipelines. These tools can re-hash deployed model graphs, compare them to stored baselines, and flag discrepancies for human review. Because ShadowLogic introduces identifiable structural changes (e.g., added constants and control-flow nodes), these diffs provide a low-cost early-warning signal of tampering.
\end{enumerate}

Beyond immediate mitigations, future work should focus on formal verification tools for computational graphs, capable of certifying functional equivalence before and after export. Integrating these techniques into ML supply-chain practices would help establish secure provenance for deployed models.
Ultimately, defending against graph-level backdoors will require treating exported model files as critical software artifacts with mandatory signature enforcement and public integrity records.
\section{Conclusion}
We demonstrated that small, targeted modifications to the computational graph of an existing white-box LLM can introduce a persistent and covert backdoor without altering model weights or retraining. ShadowLogic exposes a new and practical attack surface that arises from the absence of integrity controls in model exchange formats such as ONNX.
Beyond the uncensoring attack presented here, this method generalizes to other representational edits, both benign and malicious, underscoring the urgent need for verifiable model provenance.
Future work should pursue standardized graph-hashing frameworks and shared hash registries to ensure that LLMs distributed across research and enterprise environments can be trusted to behave as intended.
\acks{This work was supported by HiddenLayer, Inc.}

\bibliography{jmlr-sample}

\appendix

\end{document}